\documentclass[aps,pra,reprint,superscriptaddress]{revtex4-1}

\usepackage{graphicx} 
\usepackage{graphics} 
\usepackage{amsfonts} 
\usepackage{amssymb} %
\usepackage{color} 
\usepackage{epstopdf}

\begin{document}

\title{Short-pulse cross-phase modulation in an electromagnetically-induced-transparency medium
}

\author{Amir \surname{Feizpour}}
 \email{feizpour@physics.utoronto.ca}
\author{Greg \surname{Dmochowski}}
 \affiliation{Centre for Quantum Information and Quantum Control and Institute for Optical Sciences, Department
of Physics, University of Toronto, 60 St. George Street, Toronto, Ontario,
Canada M5S 1A7}
\author{Aephraim M. \surname{Steinberg}}
 \affiliation{Centre for Quantum Information and Quantum Control and Institute for Optical Sciences, Department
of Physics, University of Toronto, 60 St. George Street, Toronto, Ontario,
Canada M5S 1A7}
\affiliation{Canadian Institute For Advanced Research, 180 Dundas St. W., Toronto Ontario, Canada M5G 1Z8}

\begin{abstract}
Electromagnetically-induced transparency (EIT) has been proposed as a way to greatly enhance cross-phase modulation, with the possibility of leading to few-photon-level optical nonlinearities. 
This enhancement grows as the transparency window width,  $\Delta_{EIT}$, is narrowed.  
Decreasing $\Delta_{EIT}$, however, increases the response time of the effect, suggesting that for pulses of a given duration, there could be a fundamental limit to the strength of the nonlinearity.  
We show that in the regimes of most practical interest - narrow EIT windows perturbed by short signal pulses- the enhancement offered by EIT is not only in the magnitude of the nonlinear phase shift but in fact also in its increased duration.
That is, for the case of signal pulses much shorter (temporally) than the inverse EIT bandwidth, the narrow window serves to prolong the effect of the passing signal pulse, 
leading to an integrated phase shift that grows linearly with $1/\Delta_{EIT}$ even though the peak phase shift may saturate;
the continued growth of the integrated phase shift improves the detectability of the phase shift, in principle without bound.
For many purposes, it is this detectability which is of interest, more than the absolute magnitude of the peak phase shift.
We present analytical expressions based on a linear time-invariant model that accounts for the temporal behavior of the cross-phase modulation for several parameter ranges of interest. 
We conclude that in order to optimize the detectability of the EIT-based cross-phase shift, one should use the narrowest possible EIT window, and a signal pulse that is as broadband as the excited state linewidth and detuned by half a linewidth. 
\end{abstract}

\maketitle

\section{Introduction}
\label{sec_intro}

While photonic qubits are ideal candidates for quantum information storage and transmission, an efficient and scalable method for processing optical quantum information has yet to be demonstrated.  
The weakly interacting nature of light, which makes photonic qubits robust against decoherence, also renders photons poor candidates for information processing since (nonlinear) interactions are at the heart of logic gate operations.  

A large enough optical nonlinearity at the quantum level can pave the way for numerous applications, including low-light-level switching \cite{HarrisSwitching1998}, quantum non-demolition measurements \cite{ImotoQND}, quantum teleportation \cite{vitali2000complete}, and quantum logic gates \cite{munro2005weak}.
However, naturally occurring nonlinear optical coefficients are insufficient for these applications. 
Several different approaches have been taken to tackle the problem of very weak nonlinearities, including the use of photonic crystal fibres \cite{matsuda2009observation}, Rydberg atoms \cite{peyronel2012quantum,firstenberg2013attractive,Chen16082013,PhysRevLett.112.073901}, atoms in hollow-core fibers \cite{venkataraman2013phase}, single atoms coupled to microresonators \cite{PhysRevLett.111.193601} and even proposals to amplify the magnitude of existing nonlinear optical effects \cite{feizpour2011prl}.
Schmidt and Imamoglu proposed a scheme \cite{schmidt1996giant} based on electromagnetically induced transparency (EIT) \cite{fleischhauer2005electromagnetically} which allowed for ``giant'', resonantly-enhanced optical nonlinearities while simultaneously eliminating absorption.
While offering an orders-of-magnitude increase in interaction strength, which scales inversely with the transparency window width, this work was based on a single-mode treatment and did not consider practical details of the effect in the presence of pulsed light fields.
  
There have been several multi-mode treatments of EIT, which examine the transients due to switching on optical fields \cite{chen1998observation,li1995transient} as well as of sudden changes in two-photon (Raman) resonance \cite{godone2002rabi, park2004transient}.  
In addition, the transient properties of the associated nonlinearities, both absorptive  (photon switching) \cite{chen2004transient, shen2004influence} and dispersive (cross-Kerr effect) \cite{schmidt1998high, deng2001electromagnetically, sinclair2009time, pack2006transients, pack2007transients} have since been investigated.  
In particular, it was found that the rise time of the cross-phase modulation, that is, the time required for the phase of the probe field to reach its new steady state value in response to a step-function signal field, is inversely proportional to the EIT window width, $\Delta_{EIT}$.  
While narrow EIT windows provide a larger steady-state phase shift, more time is needed to reach this steady state.
Therefore, any attempt to increase the strength of the interaction by narrowing the EIT bandwidth would increase the response time of the nonlinear medium.  
It has since been suggested \cite{pack2007transients, sinclair2009time} that there is an inherent limitation to EIT-enhanced cross-phase modulation schemes. 
In particular, for a given signal-pulse duration, $\tau_s$, there might be a minimum tolerable window width, and hence a maximum attainable nonlinearity. 
This paper investigates whether the enhancement offered by the original proposal persists for experimentally realistic conditions which call for broadband signal pulses and narrow EIT windows.

Early schemes for optical quantum information processing required very large (on the order of $\pi$) cross-phase shifts (XPS) \cite{milburn1989fredkin}. 
As this has proven to be experimentally out of reach in single-pass geometries so far, more recent proposals have replaced the need for such large phase shifts with the less demanding requirement of any cross-phase shift detectable on a single shot \cite{munro2005weak}. 
In order to improve the detectability of the phase shift, one usually integrates the effect over its duration. 
Therefore, the integrated phase shift replaces the peak phase shift as the important figure of merit in cross-phase modulation schemes, for use as a quantum logic gate.  

Here we show that in the regime of narrow transparency windows perturbed by short signal pulses, the peak cross-phase shift saturates without shrinking and the duration of the effect grows as the window becomes narrower.
While the rise time of the EIT-enhanced XPS is determined by the signal pulse, its fall is given by the inverse EIT window width, resulting in an integrated phase shift that continues to scale inversely with the window width even for $\Delta_{EIT} \ll 1/\tau_S$.
Furthermore, we show that the dynamics of these cross-phase shifts can be understood in terms of a linear time-invariant (LTI) model.
The intensity of the signal field and the phase of the probe field can be thought of as the ``drive" and ``response" of a linear system, respectively. 
Analytical expressions based on an LTI system response accurately model the behavior of the nonlinear interaction in most regimes of interest. 

We begin by introducing in section \ref{subsec_MBE} the rigorous mathematical approach (based on the Maxwell-Bloch equations) used to study this light-matter interaction.  
Section \ref{subsec_LTI} outlines the alternative (LTI) model.
The results of these two approaches are discussed in section \ref{sec_results}, where we describe the behavior of EIT-enhanced XPS with pulsed signal fields including its dependence on various parameters of interest such as the transparency window width and the signal pulse duration.  
Finally in section \ref{subsec_OD} we discuss how propagation in an optically thick medium affects EIT-enhanced cross-phase modulation.
Throughout, we compare the predictions of an LTI model and the numerical solutions of the complete system density matrix and discuss the range of validity of such a model.

\section{Model}
\label{sec_model}

\begin{figure}[t]
  \centering %
  \def\svgwidth{\columnwidth}
  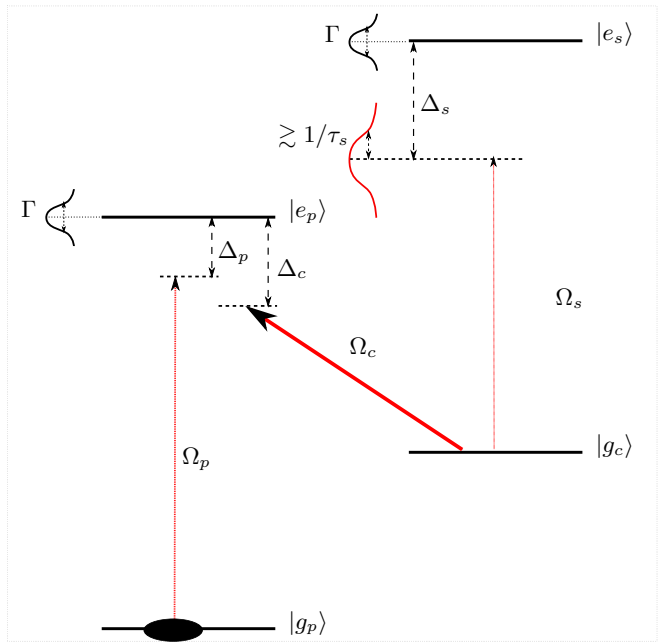
  \caption{Level structure for the simplest EIT-enhanced cross-Kerr effect, the so-called N-scheme. Here, $\Omega_p$ and $\Omega_c$ are the Rabi frequencies of the (cw) probe and coupling fields; $\Omega_s$ is the peak Rabi frequency of the signal field, which is a Gaussian pulse with length of $\tau_s$; $\Gamma$ is the excited state decay rate and $\gamma$ is the ground-state dephasing rate.
  \label{fig_LvlScheme}}
\end{figure}

Consider the level scheme shown in figure \ref{fig_LvlScheme}, in which continuous wave (cw) in-phase probe and coupling fields form a three-level Lambda system.  
If the two-photon resonance condition is satisfied, i.e. $\delta = \Delta_p - \Delta_c = 0$, and the coupling field is strong enough, $\Omega_c^2 \gg \Gamma\gamma$, then destructive interference of multiple excitation pathways causes the medium to become transparent to the probe light.  
That is, the interaction of the probe and coupling fields with the medium results in new atomic eigenstates, one of which (the so-called dark state) is decoupled from the optical fields.
Atomic population is pumped into this dark state, where it remains, at a rate of $R = \Omega_c^2 \Gamma / 2(4\Delta^2 + \Gamma^2)$, where $\Delta = (\Delta_p + \Delta_c)/2$.
The steady-state spectral width (FWHM) of the EIT window is determined by this pumping rate and the ground state dephasing rate according to $\Delta_{EIT}=2(R+\gamma)$  \cite{pack2006transients}.
The presence of the signal field inside the medium completes the `N-scheme', serving to perturb the ground-state coherence created by the Lambda system in two ways:
first, the scattering of the signal photons from the excited state $|e_s\rangle$ dephases the ground-state coherence at the rate of $\Omega_s^2 \Gamma / 4\Delta_s^2$; 
second, the Stark shift caused by the signal pulse, $\Delta_{AC}=\Omega_s^2/4\Delta_s$, detunes the system out of two-photon resonance and causes the probe field to experience a different refractive index, thereby acquiring a cross-phase shift.
The signal detuning can be made large enough compared to the excited state linewidth and the bandwidth of the signal pulse that the first contribution is negligible and only the Stark shift perturbs the system significantly. 
If this Stark shift, $\Delta_{AC}$, is smaller than the EIT window width, $\Delta_{EIT}$, then the phase shift that the probe experiences is linear in $\Delta_{AC}$ and, in turn, linear in the intensity of the signal field, $|\Omega_s|^2$.
This is the regime in which the nonlinear interaction between the signal and the probe can be considered a cross-Kerr effect.

\subsection{Maxwell-Bloch Model}
\label{subsec_MBE}
The Hamiltonian describing the interactions of figure \ref{fig_LvlScheme} (in a rotating frame and using the rotating wave approximation) is

\begin{equation}
H = \frac{\hbar}{2}\left(
\begin{array}{cccc}
0 & 0 & \Omega_p & 0 \\
0 & 2\delta & \Omega_c & \Omega_s\\
\Omega_p^* & \Omega_c^* & 2\Delta_p & 0 \\
0 & \Omega_s^* & 0 & 2(\Delta_s+\delta)
\end{array}
\right)
\end{equation}

\noindent where $\Omega_i = -\vec{\mu}\cdot \vec{E_i}/\hbar$ is the Rabi frequency and $E_i$ is the electric field for $i=p,c,s$; $\vec{\mu}$ is the matrix element of the transition.
We can find the dynamics of the system by solving the Maxwell-Bloch equations, 

\begin{widetext}
\begin{eqnarray}
\partial_t \Omega_p + c\partial_z \Omega_p & = & i g N(z) S_p(z,t) \nonumber\\
\partial_t \Omega_c + c\partial_z \Omega_c & = & i g N(z) S_c(z,t) \nonumber\\
\partial_t \Omega_s + c\partial_z \Omega_s & = & i g N(z) S_s(z,t) \nonumber\\
\partial_t S_p & = & (i \Delta_p - \Gamma/2)S_p(z,t) +i \frac{1}{2}\Omega_p(z,t) + i \frac{1}{2}\Omega_c(z,t) S_{gg}(z,t) \nonumber\\
\partial_t S_s & = & (i \Delta_s - \Gamma/2) S_s(z,t) - i \frac{1}{2}\Omega_c(z,t) S_{ee}(z,t) \nonumber\\
\partial_t S_c & = & (i \Delta_c - \Gamma/2) S_c(z,t) - i \frac{1}{2}\Omega_s(z,t) S_{ee}^*(z,t) + i \frac{1}{2}\Omega_p(z,t)  S_{gg}^*(z,t) \nonumber\\
\partial_t S_{gg} & = & (i \delta - \gamma )S_{gg}(z,t) + i \frac{1}{2}\Omega_c^*(z,t)  S_p(z,t) - i \frac{1}{2}\Omega_p(z,t) S_c^*(z,t) + i \frac{1}{2}\Omega_s^*(z,t) S_{ge}(z,t) \nonumber\\
\partial_t S_{ee} & = & \left(i (\Delta_s - \Delta_c) - \Gamma/2\right) S_{ee}(z,t) + i \frac{1}{2}\Omega_s(z,t) S_c^*(z,t) - i \frac{1}{2}\Omega_p^*(z,t)  S_{ge}(z,t) - i \frac{1}{2}\Omega_c^*(z,t)  S_{s}(z,t) \nonumber\\
\partial_t S_{ge} & = & \left(i (\Delta_s + \Delta_p - \Delta_c) - \Gamma/2 \right) S_{ge}(z,t) - i \frac{1}{2}\Omega_p(z,t) S_{ee}(z,t) + i \frac{1}{2}\Omega_s(z,t) S_{gg}(z,t)
\label{eq_EOM}
\end{eqnarray}
\end{widetext}

\noindent which encapsulate the dynamics of both the atomic system and the electromagnetic fields.
In equations \ref{eq_EOM}, $c$ is the speed of light; $N(z)$ is the atom density; $S_p=Tr(\rho |g_p\rangle \langle e_p |)$, $S_c=Tr(\rho |g_c\rangle \langle e_p |)$ and $S_s=Tr(\rho |g_c\rangle \langle e_s |)$ are the probe, coupling and signal transition coherences; $S_{gg}=Tr(\rho |g_p\rangle \langle g_c |)$, $S_{ee}=Tr(\rho |e_p\rangle \langle e_s |)$ and $S_{ge}=Tr(\rho |g_p\rangle \langle e_s |)$ are the coherences between the two ground-states, between the two excited states, and between the probe ground-state and the signal excited state, respectively; 
$\rho$ is the atomic density matrix;
and $g = \omega_0 \mu^2 / \epsilon_0 \hbar$ is the light-matter coupling constant, where $\omega_0$ is the center frequency of the electromagnetic field. 
For the purposes of this paper $\omega_0$ and $\mu$ are taken to be constants and equal for all transitions.  
In deriving the above equations of motion, it is assumed that all optical fields are weak enough that the population remains completely in the probe ground-state, $|g_p\rangle$. 
Therefore, to first order in electric fields, the equations of motion for populations can be neglected. 
We assume a Gaussian distribution for atom density and set both one- and two-photon detunings to zero, $\Delta_p = 0$ and $\Delta_c - \Delta_p = 0$, respectively.
In addition, the probe and the coupling fields are assumed to be continuous-wave (pulses with durations much longer than the simulation time) while the signal pulse is taken to be Gaussian with rms duration of $\tau_s$.
Note that the optical density of a transition is given by $d_0 = (2g/c\Gamma) \int N(z) dz = \sigma_{at} \int N(z) dz$ where $\sigma_{at}$ is the interaction cross section. 

The equations of motion, eq. \ref{eq_EOM}, can be solved using approximate analytical methods \cite{pack2006transients} or numerical techniques. 
We take the latter route, using a first-order difference method to discretize the spatial coordinate and then the 4th-order Runge-Kutta method to take the time integral, which yields the solution to the density matrix of the combined light-matter system for different sets of parameter choice.  
First, however, we present an alternate and simpler approach to modelling the dynamics as a linear time-invariant (LTI) system.  
The results of section \ref{sec_results} compare and contrast these two approaches.

\subsection{Linear Time-Invariant Model}
\label{subsec_LTI}

Here we present a model for the dynamics of the cross-Kerr interaction, which abstracts the underlying nonlinearities and treats the probe phase as a linear time-invariant ``system'' whose behavior is affected by an independent, potentially time-varying, ``driving'' signal field intensity.
The impulse response characterizing this linear system may be obtained by direct differentiation of the system's step-response.  
This step-response is precisely what has been reported in previous transient studies of EIT-enhanced XPS \cite{pack2006transients}. 
There it was shown that, when the Stark shift is smaller than the EIT window width, the rise time of the XPS is $\tau = (1+d/4)/(R+\gamma)$, where $d = d_0 R / (R + \gamma)$ is the depth of the transparency (the difference of the optical density seen by the probe on resonance without and with a resonant coupling beam).
We, therefore, take the step-response, $\mathcal{S}(t)$, to have an exponential shape, 

\begin{equation}
\mathcal{S}(t) = \frac{\phi^{ss}}{|\Omega|^2} \Theta(t) (1-\exp(-t/\tau))
\label{eq_StepResponse}
\end{equation}

\noindent
where $\phi^{ss}$ is the steady-state cross-phase shift for a weak signal field of intensity $|\Omega|^2$ and $\Theta(t)$ is the Heaviside step-function.
It is important to note that the shape of the response in an optically thick medium deviates from the exponential form.  For simplicity, we first consider optically thin media, leaving the details of optically thick samples to section \ref{subsec_OD}.  
The steady-state phase shift, $\phi^{ss}$, as predicted by single-mode and step-response treatments, is

\begin{eqnarray}
\phi^{ss} &=& \Delta_{AC} \frac{\omega_0}{2c}\int dz\left.\frac{\partial \chi_{pr}(z)}{\partial\Delta_p}\right|_{\Delta_c=0,\ \delta=0} \nonumber\\
&=& \Delta_{AC} \frac{\omega_0}{2c} \frac{4d^2}{\hbar \epsilon_0} \frac{\Omega_c^2}{(2\gamma\Gamma + \Omega_c^2)^2} \int N(z) dz \nonumber\\
&=& \Delta_{AC} d_0 \Gamma \frac{\Omega_c^2}{(2\gamma\Gamma + \Omega_c^2)^2} = \Delta_{AC}\frac{d}{\Delta_{EIT}}
\end{eqnarray}

\noindent
where $\chi_{pr}$ is the steady-state susceptibility of the probe transition \cite{fleischhauer2005electromagnetically}, $\Delta_{AC} = -|\Omega|^2 / 4\Delta_s$ is the ground-state Stark shift for $\Delta_s \gg \Gamma$, and $d/\Delta_{EIT}$ is proportional to the slope of the refractive index with respect to the detuning seen by the probe field.
The impulse response can be obtained by differentiating the above step-response,

\begin{equation}
\mathcal{I}(t) = \frac{\partial \mathcal{S}(t)}{\partial t} = \frac{\phi^{ss}}{|\Omega|^2 \tau } \Theta(t) \exp(-t/\tau)
\label{impulseResponse}
\end{equation}

Let us now investigate the behavior of this system in response to a Gaussian signal pulse. 
We describe the pulse by its time-dependent Rabi frequency,

\begin{equation}
\Omega_s(t) = \Omega_{0,s} \sqrt{\frac{1}{\tau_s\Gamma}} \exp(-t^2/4\tau_s^2).
\end{equation}

\noindent
\noindent With applications of single-photon nonlinearities in mind, we consider a fixed number of signal photons, $n_{ph}$, constraining the pulse energy,

\begin{equation}
E=\left(\sqrt{\pi}\frac{\Omega_{0,s}^2}{\Gamma^2}\frac{A}{\sigma_{at}}\right)\hbar\omega_0 = n_{ph} \hbar\omega_0,
\end{equation}

\noindent where $A$ is the transverse area of the signal pulse.
Assuming linearity, the temporal profile of the XPS is the convolution of the impulse response and the intensity profile of the signal pulse,

\begin{eqnarray}
\phi(t) &=& |\Omega_s(t)|^2 * \mathcal{I}(t) \nonumber\\
&=& \frac{\phi_0 n_{ph}}{2\tau} e^{\tau_s^2/2\tau^2} \nonumber\\
&& \times \exp(-t/\tau) \left(1 + \textrm{erf}(t/\sqrt{2}\tau_s - \tau_s/\sqrt{2}\tau)\right)
\label{phiOfTee}
\end{eqnarray}

\noindent where $\textrm{erf}(x) = 2/\sqrt{\pi} \int_0^x dx' \exp(-x'^2)$ is the error function, $*$ indicates convolution and

\begin{equation}
\phi_0 = \frac{\Gamma}{-4\Delta_s} \frac{\sigma_{at}}{A} \frac{d}{\Delta_{EIT}}
\label{eq_phi_int}
\end{equation}

\noindent
is the integrated XPS per signal photon.
The temporal profile of the XPS predicted by the LTI model, eq. \ref{phiOfTee}, suggests that there are two different timescales involved: the response time of the EIT medium, $\tau$, and the signal pulse duration, $\tau_s$.  
Initially, when $t \ll \tau$, the error function term alone dictates the temporal shape, having a timescale given by $\tau_s$.  
The rise of the phase shift always mimics the envelope of the signal pulse, irrespective of $\tau$.
For later times, however, the temporal shape of the phase shift is given by a combination of the signal pulse duration and the response time of the EIT medium.
In the limiting case of $\tau_s \gg \tau$ (when the signal pulse is much longer than the response time of the medium), the system follows the signal pulse envelope.
This corresponds to a quasi-steady-state scenario where the atomic coherences are able to follow the change in two-photon detuning arising from the signal field.
In the other extreme, when $\tau_s \ll \tau$, the phase of the probe field rises quickly due to the short signal pulse and then relaxes to its original steady-state value on a timescale given by $\tau$ alone.  
This corresponds to a short impulse perturbing the system momentarily, leaving the atomic coherences to build back up once it passes.
For intermediate cases, the phase decays on a timescale which is a combination of $\tau$ and $\tau_s$.

In addition, the integrated phase shift per photon, $\phi_0$, as predicted by the LTI model, eq. \ref{eq_phi_int}, is seen to be independent of the signal pulse duration (recall that the energy of the signal pulse was held fixed).
Importantly, the integrated phase shift scales inversely with EIT window width for pumping rates much larger than the dephasing rate, $R \gg \gamma$; peaks when $R = \gamma$; and falls off for $R\ll \gamma$. 
The only other parameters that $\phi_0$ depends on are the optical density $d_0$, the signal pulse detuning $\Delta_s$, and how tightly the signal beam is focused compared to the atomic cross section, $\sigma_{at}/A$.  
We now turn to the dynamics of EIT-enhanced XPS and show that this linear model accurately predicts the behavior obtained from a numerical solution of the complete system density matrix.

\section{Results}
\label{sec_results}

In what follows, we show how different parameters of interest modify the behavior of EIT-enhanced XPS in the presence of a pulsed signal field.  
We consider both the numerical solution of section \ref{subsec_MBE} as well as the LTI model of section \ref{subsec_LTI} and show that the latter captures the salient features of this nonlinear interaction.  
We begin by discussing the effect of the transparency window width, $\Delta_{EIT}$, on the XPS time response and the role that dephashing plays in this regard.  In section \ref{subsec_sig}, we investigate the effects of the signal pulse duration and detuning, and we conclude by discussing in section \ref{subsec_OD} how an optically thick medium alters these dynamics.

\subsection{Dependence on EIT medium properties}
\label{subsec_EIT}

\begin{figure}[t]
  \centering %
  \includegraphics[width = \columnwidth]{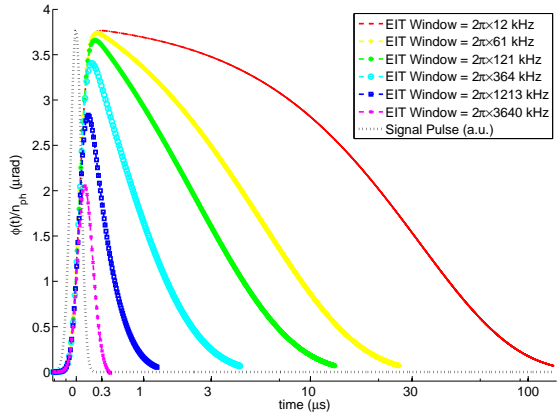}
  \caption{Time dependence of the per-photon cross-phase shift for a variety of EIT window widths. The linear scaling of the peak XPS versus EIT window width breaks down once the response time of the EIT medium becomes comparable to or larger than the signal pulse duration. However, narrower window widths produce longer tails. Simulation parameters: $\Gamma=2\pi\times 6$MHz, $\tau_s = 1/2\sqrt{2}\pi \times 2000$ kHz$^{-1}$, $n_{ph} = 100$, $d_0 = 1$, $\Delta_p = 0$, $\Delta_c = 0$, $\Delta_s = - 10\Gamma$, $\sigma_{at} = 1.2\times 10^{-13}\ m^2$, $\Omega_{0,p} = 0.003 \Gamma$, $\gamma  =1\times 10^{-5} \Gamma$, beam waist is $10\ \mu$m and the wavelength is 780.24 nm. The atomic cloud has a Gaussian spatial distribution.
  \label{fig_TimeTrace_vs_Window}}
\end{figure}

\begin{figure}[t]
  \centering %
  \includegraphics[width = \columnwidth]{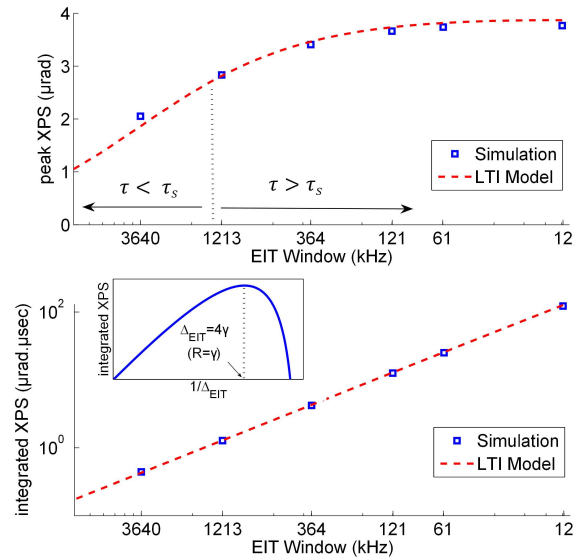}
  \caption{Peak (top) and integrated (bottom) XPS per photon as extracted from figure \ref{fig_TimeTrace_vs_Window}. 
  The peak XPS scales inversely with EIT window only when the response time of the EIT medium is shorter than the signal pulse duration while the integrated phase shift grows inversely with window width owing to the longer tails that arise from narrower EIT windows. 
  Squares correspond to simulation results and dashed lines show the prediction of the LTI model presented in section \ref{subsec_LTI}. 
  For window widths comparable to the natural linewidth of the transition the EIT medium response includes oscillations that are not included in the LTI impulse response, resulting in a small discrepancy between the two approaches. 
  Also, the linear scaling of the integrated phase shift can be interrupted if the pumping and dephasing rates become comparable (inset).
  \label{fig_pk_vs_Window}}
\end{figure}

\begin{figure}[t]
  \centering %
  \includegraphics[width = \columnwidth]{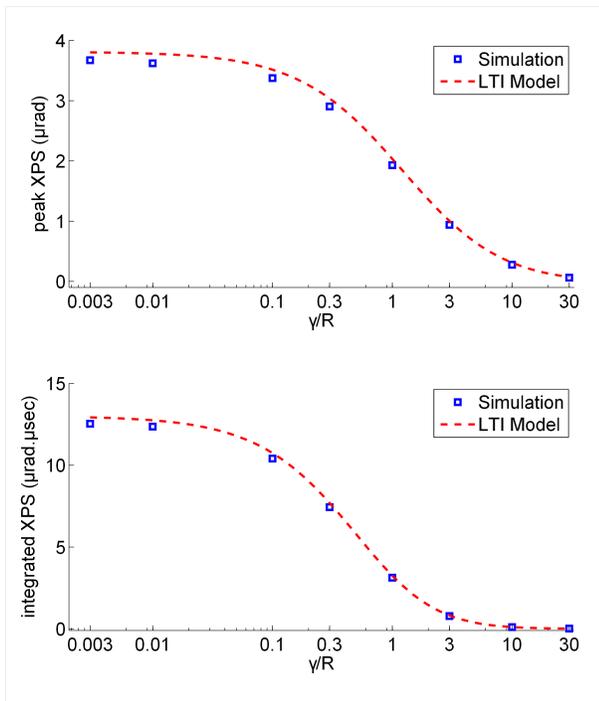}
  \caption{Peak (top) and integrated (bottom) XPS per photon for various ground-state dephasing rates, $\gamma$. As the dephasing rate increases, both peak and integrated XPS decrease due to the degradation of the EIT window. Peak XPS falls to nearly half of its ideal value when the dephasing rate becomes equal to the pumping rate, $R$. Squares correspond to simulation results while the dashed lines show the prediction of the LTI system response. For this simulation $R = 0.01\Gamma$, $\tau_s = (0.6\Gamma)^{-1}$ and the rest of parameters are the same as in figure \ref{fig_TimeTrace_vs_Window}. Note that EIT window width is $2(R+\gamma)$. 
  \label{fig_pk_vs_gam}}
\end{figure}

We first address how the width of the transparency window affects the dynamics of the EIT-enhanced XPS. 
In the original single-mode treatment, the size of the nonlinear phase shift increased indefinitely as the EIT window was narrowed.
In the subsequent multi-mode, step-response analysis, the steady-state phase shift behaved similarly but this steady state took longer to be established for narrower transparency windows.
Figure \ref{fig_TimeTrace_vs_Window} shows the temporal profile of the XPS experienced by a probe field in response to a Gaussian signal pulse for a variety of EIT window widths, as obtained by numerical simulation of equation \ref{eq_EOM}.
It is immediately evident that the rise time of the nonlinear phase shift is independent of the EIT window width, mimicking instead the rise of the signal pulse; also, as the window width narrows, the effect of the signal pulse on the probe field is prolonged.
For narrower EIT windows, more time is required for the probe phase to return to its original steady-state value. 
In many practical applications of the EIT-enhanced cross-Kerr effect, this elongated tail permits a longer integration time and, hence, improved signal-to-noise.

Figure \ref{fig_pk_vs_Window} shows the peak and integrated phase shifts extracted from figure \ref{fig_TimeTrace_vs_Window} (squares) as well as those predicted from the LTI model of section \ref{subsec_LTI} (dashed line). 
Immediately evident is the good agreement between these two different approaches.
In both cases, we see that the peak phase shift scales linearly with $1/\Delta_{EIT}$ only when the EIT window is wide enough that $\tau \ll \tau_s$, i.e. when the response time is shorter than the signal pulse duration; once the window becomes narrower this linear scaling is disrupted, eventually plateauing for $\tau \gg \tau_s$.
In fact, the peak phase shift changes by a mere factor of two for a window width variation that spans two orders of magnitude.
Although the steady-state phase continues to grow with decreasing $\Delta_{EIT}$, the time needed to reach this steady state also grows while the interaction time (signal pulse duration) is held constant here.
Therefore, once $\Delta_{EIT}$ is sufficiently narrow, decreasing the window width further does not help with increasing the peak phase shift, which accounts for the plateau seen in figure \ref{fig_pk_vs_Window}.
On the other hand, figure \ref{fig_pk_vs_Window} also shows that the integrated phase continues to scale inversely with the EIT window width irrespective of the medium response time and the signal pulse duration.  
We are, therefore, led to conclude that the slow dynamics, far from degrading the effect, can still lead to an enhanced integrated cross-phase shift that could be exploited to obtain better signal-to-noise ratio when detecting an EIT-based cross-phase shift, even when the peak phase shift saturates.

We see that the integrated phase shift scales as $1/\Delta_{EIT}$ and this scaling is interrupted only by the ground-state dephasing rate, $\gamma$, which has only technical but no fundamental limit.  
This dephasing limits the maximum depth of transparency, $d = d_0R/(R+\gamma)$, as well as the minimum attainable EIT window width, $2(R+\gamma)$.
These two quantities correspond to the rise and run, respectively, of the refractive index profile experienced by the probe field.
Figure \ref{fig_pk_vs_gam} shows the peak and integrated cross-phase shifts for various values of $\gamma$ and a fixed pumping rate, $R$. 
The peak XPS falls by a factor of two at $\gamma = R$ while the integrated XPS does so at a value of $\gamma$ smaller than $R$ since it is affected by both the refractive index slope and the shortened tail.

\subsection{Dependence on signal pulse}
\label{subsec_sig}

\begin{figure}[t]
  \centering %
  \includegraphics[width = \columnwidth]{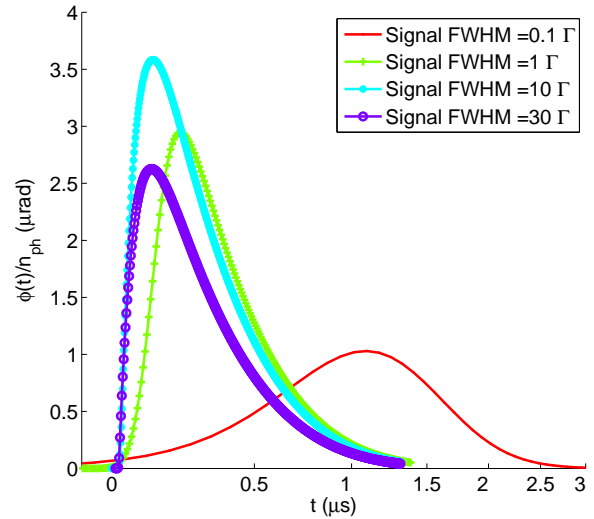}
  \caption{Time response of XPS (per photon) for various signal pulse bandwidths. 
  The linear scaling of the peak phase shift with signal pulse bandwidth breaks down when this bandwidth becomes comparable to or larger than the EIT window width. 
  Once the bandwidth of the signal pulse becomes comparable to its detuning, $\Delta_s = -10\Gamma$, the peak XPS stops growing and starts to fall. Simulation parameters: $\Delta_{EIT} = 0.2\Gamma$ and the rest of parameters are the same as in figure \ref{fig_TimeTrace_vs_Window}.
  \label{fig_TimeTrace_vs_sig}}
\end{figure}

\begin{figure}[t]
  \centering %
  \includegraphics[width = \columnwidth]{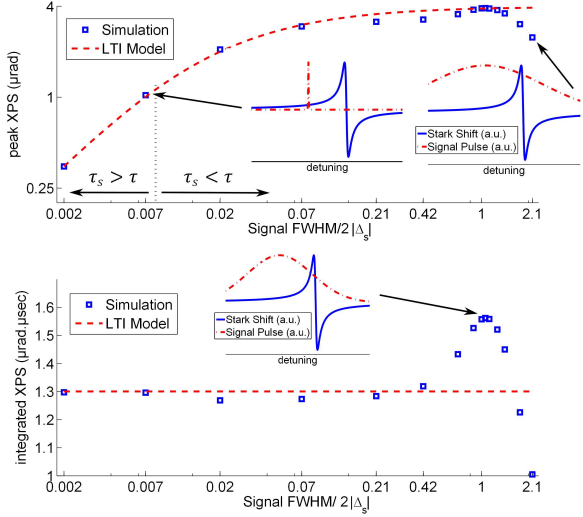}
  \caption{Peak (top) and integrated (bottom) XPS per photon as a function of signal field bandwidth (normalized to central detuning) as extracted from figure \ref{fig_TimeTrace_vs_sig}. Initially, increasing the pulse bandwidth causes the peak XPS to grow proportionately due to the higher pulse intensity. However, once the pulse bandwidth becomes larger than the EIT window width, the peak XPS stops growing, similar to the behavior seen in figure \ref{fig_pk_vs_Window}. The maximum integrated XPS occurs when the pulse half-width at half-maximum of the intensity is equal to the detuning. The insets show the Fourier transform of the signal pulse intensity (red dashed) along with the frequency dependence of the ac-Stark shift (blue solid) as a function of detuning from the excited state. For very broadband pulses, there is a discrepancy between the result of the LTI model and the numerical solution as explained in the text. 
  \label{fig_pk_vs_sig}}
\end{figure}

\begin{figure}[t]
  \centering %
  \includegraphics[width = \columnwidth]{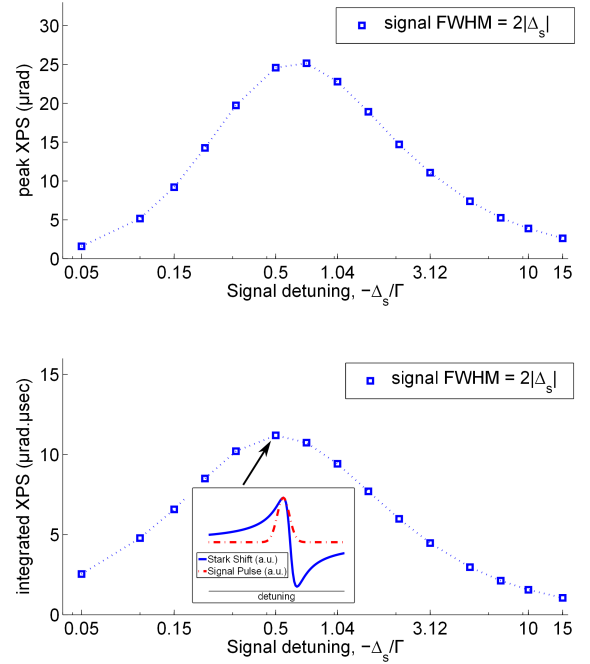}
  \caption{Peak (top) and integrated (bottom) XPS per photon for various signal detunings, $\Delta_S$, when the half-width at half-maximum of the signal pulse bandwidth is set equal to the detuning. The squares show simulation results while the dotted line is a guide for the eye. Both peak and integrated XPS have maxima close to $\Delta_s = \Gamma/2$. 
  The inset shows the Fourier transform of the signal pulse intensity (red dashed) along with the frequency dependence of the ac-Stark shift (blue solid) as a function of detuning from the excited state. 
  \label{fig_pk_vs_del}}
\end{figure}

So far the only assumption we have made about the frequency content of the signal pulse was that its bandwidth was small compared to the signal pulse detuning.
In this section we study how changing this frequency content can result in the modification of the behavior of the EIT-enhanced cross-phase shift.
For simplicity we assume that the signal pulse is transform-limited: that is, that its bandwidth is proportional to $1/\tau_s$. 
Increasing the bandwidth, therefore, corresponds to a temporally shorter pulse.  
Since the Kerr effect depends linearly on the signal field intensity, one would expect to be able to maximize the cross-phase shift, for a given pulse energy, by making the pulse as short, and therefore as intense, as possible.
However, in the case that the spatial extent of the signal pulse is larger than the atomic medium, a shorter pulse yields a shorter interaction time and this must be weighed against the larger intensity due to broadening the signal bandwidth (i.e. decreasing $\tau_s$).

Figure \ref{fig_TimeTrace_vs_sig} shows the temporal profile of the cross-phase shift for different signal pulse bandwidths for a constant pumping rate of $R=0.1\Gamma$. 
We find that when $\tau_s \gg \tau$, the cross-phase shift replicates the temporal profile of the signal pulse but the peak phase shift is relatively small due to the low intensity signal pulse.
As one broadens the bandwidth of the pulse, the peak intensity and therefore the peak phase shift increase.
However, this increase in peak phase shift with signal intensity is seen to saturate and even reverse once $\tau_s$ becomes sufficiently small.
Figure \ref{fig_pk_vs_sig} plots the peak and integrated cross-phase shift against signal pulse bandwidth normalized to its detuning, $\Delta_s$.
For pulse bandwidths narrower than the EIT window the peak phase shift scales linearly with signal bandwidth (and therefore linearly with intensity) as expected from single-mode or step-response treatments.  
However, once the signal bandwidth exceeds the EIT window width, the scaling begins to flatten out.
This saturation is a consequence of the tradeoff between shorter interaction time and higher peak intensity of the signal pulse.
Once the signal pulse has a bandwidth wider than the EIT window then it exits the medium before the cross-phase shift reaches its peak value.  
Increasing the bandwidth any more does not lead to a larger peak phase shift. 
The integrated cross-phase shift remains flat throughout all of this due to the fact that we have held the energy of pulse and the window width constant.

Once the bandwidth of the signal pulse grows to be comparable to its detuning, the variation of the signal pulse amplitude versus frequency becomes important.
The response function used in section \ref{subsec_LTI} does not take that frequency content into account and therefore fails to predict the behavior of the system properly.  
We can, however, qualitatively understand the behavior of XPS due to broadband pulses by recalling that the frequency dependence of the Stark effect resembles a refractive index profile.  
That is, it is an odd function passing though zero on resonance, with extremes $\Gamma/2$ away on either side of resonance and scaling inversely with detuning away from resonance.
Therefore, for a given signal pulse detuning, as its bandwidth is broadened, a point will be reached when frequency components begin to encroach on the peak of the Stark profile, leading to a larger cross-phase shift.
However, as the bandwidth is broadened further, this increase is quickly reversed as frequency components begin to cross over to the other side of the resonance addressed by this signal field.  
These frequency components then contribute strongly to the Stark shift but with opposite sign, yielding a smaller net phase shift.
The optimum phase shift is obtained when the signal half-width at half maximum (HWHM), $\sqrt{\log 2} / \sqrt{2}\tau_s$, is equal to the signal detuning, $\Delta_s$. 

It is interesting to see how the XPS behaves as a function of signal detuning when $\Delta_s \tau_s$ is held constant at the value of $\sqrt{\log 2} / \sqrt{2}$.
Figure \ref{fig_pk_vs_del} shows the peak and integrated XPS for the case when the signal HWHM is set equal to the detuning and then the two are varied simultaneously.
It can be seen that the largest optimum phase shift is achieved close to $\Delta_s = \sqrt{\log 2} / \sqrt{2}\tau_s = \Gamma/2$.
For this choice of detuning and signal bandwidth the center of the pulse (in frequency space) coincides with the peak of the Stark shift frequency profile and its width covers those parts with the largest positive shift without spilling over onto the other side of the resonance (inset of figure \ref{fig_pk_vs_del}).

\subsection{Propagation in an optically thick medium}
\label{subsec_OD}

\begin{figure}[t]
  \centering %
  \includegraphics[width = \columnwidth]{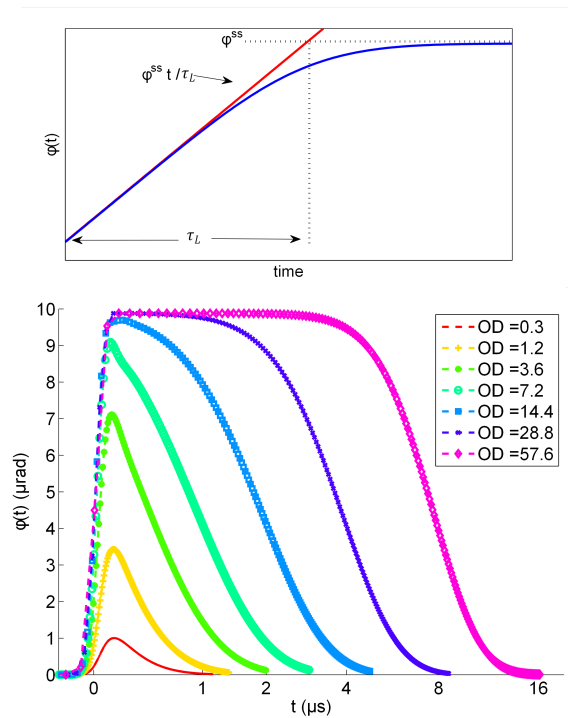}
  \caption{XPS due to a step-function signal field (top) and time dependence of pulsed XPS (per photon) for different optical densities (bottom). As the OD, $d_0$, increases the peak XPS begins to grow but eventually saturates due to the group velocity mismatch between the signal and the probe. However, larger values of optical density result in longer-lasting phase shifts; the temporal extent of the flat region of the transient is determined by the duration of the probe that is compressed in the medium, $\tau_L$, when the signal pulse passes through the medium at group velocity, $c$. Simulation parameters: $R = 0.1\Gamma$,  $\tau_s = (0.6\Gamma)^{-1}$ and all others as in figure \ref{fig_TimeTrace_vs_Window}. 
  \label{fig_TimeTrace_vs_OD}}
\end{figure}

\begin{figure}[t]
  \centering %
  \includegraphics[width = \columnwidth]{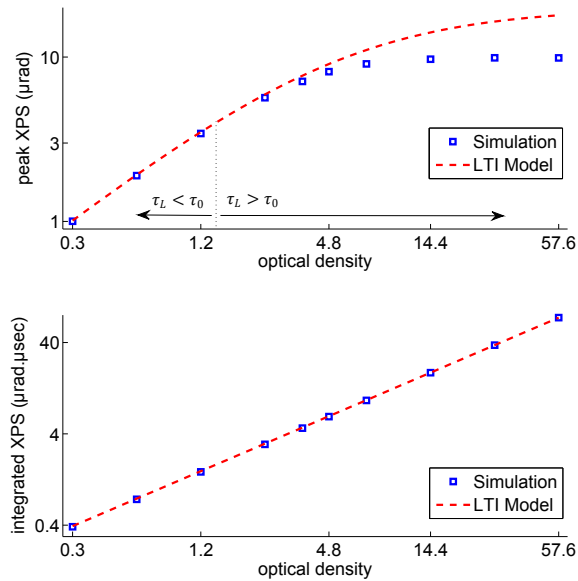}
  \caption{Peak (top) and integrated (bottom) XPS per photon versus optical density, $d_0$, as extracted from figure \ref{fig_TimeTrace_vs_OD}. Squares correspond to simulation results while the dashed lines are predictions of an LTI model. 
  The response function adopted in equation \ref{impulseResponse} only partially accounts for the propagation effects (through the dependence of the EIT medium response time, $\tau$, on OD); 
  however, this is not sufficient to model the behavior of the system at high optical densities. 
  It is important to note that the response of the system is still linear at high optical densities and a proper impulse response can account completely for the saturation effect. 
  The integrated XPS increases linearly with OD and an LTI model agrees very well with the simulation results. 
  $\tau_0$ is the response time of the EIT medium in the limit of vanishing optical density.
  \label{fig_pk_vs_OD}}
\end{figure}

Thus far, we have neglected the effects that an optically thick medium would have on the dynamics of EIT-enhanced XPS.
Steady state analysis predicts that the cross-phase shift scales linearly with the optical density (OD) and so it is of interest to see how the dynamics are affected by exploiting higher optical densities.  
Particularly in the presence of EIT, which eliminates linear absorption, higher OD increases the nonlinear interaction with no detrimental effects arising from absorption. However, for pulsed fields increasing the optical thickness of the sample increases the difference in group velocities between the probe and the signal fields.

For a sufficiently high optical density, the transit time of the probe beam through the sample becomes longer than the temporal duration of the signal pulse.  
In this case, there will be portions of the probe field inside the medium which experience the entire signal pulse as it passes through and, therefore, these portions acquire the maximum phase shift possible.  
The temporal length of this portion of the probe is equal to its group delay, $\tau_L = L/v_g = d_0 (R - 2\gamma^2/\Gamma)/2(\gamma + R)^2$ where $L$ is the length of the medium and $v_g$ is the group velocity of the probe.
This is reflected in figure \ref{fig_TimeTrace_vs_OD}, where we plot the temporal profiles of the cross-phase shifts for a variety of optical densities.  
We see that for high OD, the peak height of the phase shift plateaus but the duration of this peak cross-phase shift continues to grow as the OD is increased.   
The net effect, as shown in figure \ref{fig_pk_vs_OD}, is such that while the peak phase shift saturates, the integrated phase shift scales linearly with optical density.  

To determine this saturation value of the peak XPS, it is instructive to consider the response of the system to a step signal, see figure \ref{fig_TimeTrace_vs_OD} (top), which includes a linear rise with time-scale $\tau_L$, followed by an exponential approach to the steady-state value.
The slope of the rise, shown by the red line in figure \ref{fig_TimeTrace_vs_OD} (top), is equal to $\frac{\phi^{ss}}{\tau_L}$. 
Since the impulse response is the derivative of the step response, this slope determines the maximum achievable XPS for pulsed signal in the presence of high optical density,

\begin{equation}
\phi_{max} = \frac{\phi^{ss}}{\tau_L} \int dt |\Omega_s(t)|^2 = -\frac{\Gamma}{4\Delta_s}\frac{\sigma_{at}}{A}
\end{equation}

\noindent This is similar to the limit found by Harris and Hau due to group velocity mismatch in N-scheme \cite{PRL_HarrisHau}.
Unlike the case of the response to a step signal, where the propagation effects show up in the rise time of the nonlinear phase shift \cite{pack2007transients}, the response to a pulsed signal has a  rise time determined by the signal pulse and the propagation effects only result in the saturation of the peak XPS in the time response.

We also see that while the integrated phase shift is well modelled by our LTI approach, the peak phase shift is under-estimated for sufficiently high optical densities.  
This does not result from a breakdown of the linearity but rather because the response function assumed in section \ref{subsec_LTI} did not account for such propagation effects.  
In an optically thick medium the effect from each thin slab of the medium takes some time, determined by the group velocity of the probe and the length of the medium, to reach the observer.  
Therefore, the exponential rise assumed in equation \ref{eq_StepResponse} does not capture the additional group delay effects present in media with high optical densities.

\section{Summary}
We studied the behavior of EIT-enhanced XPS for pulsed signals in the N-scheme, and showed how different parameters, such as EIT window width, pulse bandwidth, and optical thickness affect the transient behavior of the system. 
The results obtained here have important implications for quantum logic gates based on such EIT schemes; 
they also permitted us to determine the optimal pulse duration and detuning for these purposes. 
We also showed that a treatment based on linear time-invariant system response, taking the intensity of the signal as the ``drive" and the phase shift on the probe as the ``output", adequately models the transient behavior of the Kerr cross-phase shift. 

The peak value and the duration of XPS are determined by several parameters while the rise time of the effect is always dictated by the signal pulse duration.
The peak XPS scales as the inverse of the EIT window width and is linear in pulse bandwidth as long as the EIT window is broader than the pulse bandwidth.
However, for EIT windows narrower than the pulse bandwidth, even though there is no increase in the peak XPS, the effect lasts for a longer time, providing more time for detecting the phase shift and potentially improving the signal-to-noise ratio.
The peak XPS also scales linearly in optical density as long as propagation effects can be neglected.
For optical densities above $\sim2$ (assuming negligible dephasing), the group velocity mismatch of the probe and the signal starts to play a significant role in the dynamics of the response and this poses a limitation on the maximum achievable peak phase shift.
On the other hand, this group velocity mismatch causes the XPS to last longer.
In short, narrow EIT windows and high optical densities can enhance the detectability of cross-phase shift by elongating the duration of the effect.

For short signal pulses, when the bandwidth of the pulse becomes comparable to or larger than its detuning, it becomes necessary to take the frequency dependence of the Stark effect into account.
Most importantly, the components of the signal pulse closer to the transition produce a larger Stark shift and consequently a stronger XPS.
We showed, using numerical solutions, that the optimum signal bandwidth is on the order of the signal detuning.
The largest optimum XPS is achieved when both the detuning and the HWHM of the signal pulse are equal to the half-linewidth of the excited state. 

Unlike the peak cross-phase shift, which is limited by the EIT response time and propagation effects, the integrated phase shift follows the prediction of the steady-state treatment. 
This integrated phase shift, which grows linearly with OD and inversely with EIT window width, is a more relevant figure of merit for the detectability of the XPS \cite{munro2005weak}. 
The results presented here demonstrate that, contrary to earlier fears about finite response time, EIT may indeed be used to greatly enhance nonlinear phase shifts for applications such as quantum information processing.

\section*{Acknowledgement}
 This work was funded by NSERC, CIFAR, and QuantumWorks. We would like to thank Tilman Pfau for many stimulating discussions, and for first raising the possibility that the finite response time of EIT might put the kaibosh on this scheme.
 We also thank Hamidreza Kaviani for discussions about the numerical methods. 

\bibliographystyle{apsrev4-1}
\bibliography{refs}

\end{document}